\documentclass[aps,twocolumn,pra,superscriptaddress,articlepaper,nofootinbib,longbibliography,floatfix]{revtex4-2}

\usepackage{amsmath, amsthm, amssymb, amsfonts}
\usepackage{mathtools}
\usepackage{upgreek}
\usepackage{booktabs}
\usepackage{url}
\usepackage{float}
\usepackage{enumitem}
\usepackage{printlen}
\usepackage{lineno}
\usepackage{siunitx}
\usepackage{graphicx}
\usepackage{dcolumn}
\usepackage{bm}
\usepackage{txfonts}
\usepackage[dvipsnames, svgnames, table]{xcolor}
\usepackage[colorlinks=true, citecolor=Blue,linkcolor=BrickRed, urlcolor=magenta]{hyperref}
\usepackage{placeins} 
\usepackage{setspace}
\usepackage[final]{changes}
\usepackage{comment}

\begin{document}


\title{Development of a Photonically Generated Frequency Reference for Space VLBI}



\author{Hannah Tomio}
\email{hannah.tomio@unibe.ch} 
\affiliation{Department of Aeronautics and Astronautics, Massachusetts Institute of Technology, 77 Massachusetts Ave., Cambridge, MA 02139, USA}
\author{Kohei Yamamoto}
\affiliation{Center for Space Sciences and Technology, University of Maryland, Baltimore County, 1000 Hilltop Circle, Baltimore, MD 21250, USA}
\affiliation{NASA Goddard Space Flight Center, 8800 Greenbelt Road, Greenbelt, MD 20771, USA}
\affiliation{Center for Research and Exploration in Space Science and Technology, NASA/GSFC, 8800 Greenbelt Road, Greenbelt, MD 20771, USA}
\author{Charlotte Zehnder}
\affiliation{Department of Physics, University of Arizona, 1118 E. Fourth Street, Tucson, AZ 85721, USA}
\author{Guangning Yang}
\affiliation{NASA Goddard Space Flight Center, 8800 Greenbelt Road, Greenbelt, MD 20771, USA}
\author{Kenji Numata}
\affiliation{NASA Goddard Space Flight Center, 8800 Greenbelt Road, Greenbelt, MD 20771, USA}
\author{Andrew Attar}
\affiliation{Vescent Technologies, 14998 W. 6th Ave., Suite 700, Golden, CO 80401, USA}
\author{Henry Timmers}
\affiliation{Vescent Technologies, 14998 W. 6th Ave., Suite 700, Golden, CO 80401, USA}
\author{Holly Leopardi}
\affiliation{NASA Goddard Space Flight Center, 8800 Greenbelt Road, Greenbelt, MD 20771, USA}


\date{\today}

\begin{abstract}
Space-based Very Long Baseline Interferometry (VLBI) can improve both the angular and temporal resolution of images over terrestrial VLBI techniques. 
Future space VLBI (sVLBI) missions are targeting higher observation frequencies than modern, terrestrial VLBI, and therefore these missions will require higher stability frequency references than standard hydrogen masers. 
While existing frequency references for space-based applications have concentrated on long-term stability, in this effort, we present a frequency reference with high short-term stability intended for sVLBI missions. 
This reference leverages developments in cavity-stabilized lasers from the Laser Interferometer Space Antenna (LISA) mission and commercial efforts towards a robust, space-qualifiable optical frequency comb.
We describe the implementation and experimental results of this system with a path to spaceflight, which demonstrates an Allan deviation of \num{3.26d-14} at \SI{1}{\second}, \num{6.00d-15} at \SI{10}{\second}, and \num{6.73d-15} at \SI{30}{\second} for the \SI{100}{\mega\Hz} frequency reference signal, and show that this performance meets the needs of future sVLBI missions.
\end{abstract}


\maketitle

\section{Introduction}
Very Long Baseline Interferometry (VLBI) is an observational technique in astronomy that enables the high-resolution imaging of cosmic radio sources. 
Signals from these distant sources are observed at multiple, widely separated radio telescopes, then mixed to baseband, digitized, and recorded through the telescope receiver and backend electronics.
After aligning the recorded data from each telescope to a common time frame, the data can be cross-correlated for all pairs of telescopes, calibrated, and averaged in post-processing. 
From the resulting interference fringes, the complex-valued spatial Fourier components (or visibilities) of the source brightness can be retrieved.
A VLBI array samples this brightness distribution in Fourier space, also referred as the $(u,v)$ plane.
The $(u,v)$ sampling increases with the number of baselines (distance between telescopes) of differing length and orientation to the source, and with the duration of the observations.
These samples can then be combined to construct a high-resolution image.
Fundamentally, the resolution that the array can achieve is dictated by the length of the longest baseline ($D$) and the observing wavelength ($\lambda$), such that the angular resolution is $\lambda/D$.
With an observing frequency of \SI{230}{\giga\Hz} ($\lambda = 1.3$ mm), the Earth-spanning array of radio telescopes comprising the Event Horizon Telescope (EHT) has demonstrated the power of VLBI by producing images of the supermassive black holes at the centers of galaxy M87 and the Milky Way \cite{eht_m87_i_2019, eht_sga_i_2022}.

Extending terrestrial telescope arrays with a space-based segment, or even realizing a fully orbiting array, can offer several advantages and further improve imaging capability \cite{fromm_2021, palumbo_2019}. 
A space VLBI (sVLBI) mission can provide higher angular resolution by enabling baselines longer than Earth’s diameter and allowing for the use of higher observing frequencies that are limited on ground due to atmospheric absorption \cite{johnson_2024}. 
In addition, turbulence in the atmosphere reduces the coherence time, or the time interval over which coherence can be maintained, of the received signals, which limits the possible integration time. 
Avoiding atmospheric turbulence can thereby improve the VLBI signal-to-noise ratio (SNR) \cite{rioja_2012}.
A higher temporal resolution can also be achieved with an sVLBI mission.
For ground-based VLBI, the temporal resolution is limited by the rotation rate of the Earth, which is used to sample the $(u,v)$ plane.
One or more spacecraft with shorter orbital periods could provide denser $(u,v)$ sampling more quickly, enabling imaging over shorter timescales \cite{palumbo_2019}.

Previous sVLBI missions have already demonstrated the feasibility of this concept \cite{gurvits_2020}. 
However, the next generation of missions targeting higher observing frequencies and offering much higher sensitivity are currently being proposed \cite{johnson_2024, trippe_2024, gurvits_2022}. 
These missions, among others, consist of both space-space VLBI concepts, in which two or more orbiting telescopes form the interferometer fully in space, and space-ground VLBI concepts, in which one or more orbiting telescopes augment terrestrial VLBI arrays.
A space-space VLBI scheme enables very high angular resolution by forming baselines that can far exceed Earth’s diameter and observing at much higher frequencies unaffected by Earth’s atmosphere, while a space-ground VLBI scheme can offer an improved angular resolution using a long baseline, while still maintaining a dense sampling of the $(u, v)$ coverage through the Earth-based VLBI arrays.
All of these missions will require several technological advancements, including the development of highly stable, space-qualifiable frequency references.

A highly stable frequency reference is necessary at each telescope node in a VLBI array to ensure the recorded signals can be cross-correlated and still maintain coherence.
The phase stability of this reference is a dominant factor limiting the coherence time of the interferometer formed by a pair of telescopes \cite{thompson_2017}.
On Earth, however, atmospheric fluctuations also reduce the coherence time to tens of seconds for millimeter and sub-millimeter wavelength observations \cite{pesce_2024}.
These effects in turn limit the typical VLBI integration times to just a few tens of seconds \cite{eht_m87_ii_2019}.
Like terrestrial VLBI arrays, space-ground VLBI concepts still contend with an integration time limited by the timescales of atmospheric coherence over the ground site.
However, there are still constraints on the integration time, resulting from thermal noise and the speed of the spacecraft through the $(u, v)$ plane, when both nodes of the interferometer are in space \cite{palumbo_2019}.
Hence, for both space-space and space-ground VLBI concepts, short-term stability ($\tau < $ \SI{30}{\second}) is the most important clock performance metric. 

An upper bound for the required stability of the frequency reference is given by
\begin{equation}
    2\pi\nu_0\tau_c\sigma_y(\tau_c) < 1~\unit{\radian}
    \label{eqn:basic_coherence_func}
\end{equation}
where $\nu_0$ is the observation frequency (in \unit{\Hz}) and $\sigma_y(\tau)$ is the fractional frequency instability (or Allan deviation \cite{allan_1966}) of the reference, and $\tau_c$ is the coherence time.
This simple relation denotes the required Allan deviation of the reference to limit the accumulation of phase fluctuation differences between two antennas to below \SI{1}{\radian} in order to avoid decoherence effects and unambiguously detect the VLBI interference fringe \cite{rogers_1981}.

For the current, standard observing frequency of \SI{230}{\giga\Hz}, this relation implies a performance bound of $\sigma_y(\tau) \leq $ \num{6.9d-13} at \SI{1}{\second} and \num{2.3d-14} at \SI{30}{\second}.
This performance requirement can be achieved with active hydrogen maser frequency standards, which are typically utilized for ground-based radio telescopes, such as those that comprise the EHT \cite{eht_m87_ii_2019}, and have been used for previous sVLBI missions, either as on-board references \cite{belyaev_2013} or through free-space frequency transfer to a ground-based reference \cite{hirosawa_2002}. 
However, with increasing interest in shorter wavelength/higher frequency observations, both terrestrially \cite{pesce_2021} and in future sVLBI missions (targeting \SI{320}{\giga\Hz} \cite{johnson_2024}, \SI{690}{\giga\Hz} \cite{trippe_2024}, and \SI{1}{\tera\Hz} \cite{gurvits_2022}), frequency references with better stability are necessary.
For sVLBI, a lighter, more compact, and power-efficient system is also advantageous, as a space platform has limited size, weight, and power (SWaP) resources available to any given subsystem.
A frequency reference for space must also be highly robust, as the launch conditions and space environment also pose numerous challenges, including dynamic vibration and shock, extreme temperature, and radiation.

Recent development and demonstrations of frequency references specifically for space have explored a variety of microwave atomic clock technologies \cite{burt_2021, liu_2018, deng_2024, laurent_2015, cacciapuoti_2024, jaduszliwer_2021, gozzelino_2023}. 
These efforts have concentrated on long term stability ($\tau > 1000$) for applications in Global Navigation Satellite Services (GNSS), deep space navigation, and fundamental physics tests.
The use of high precision time transfer, based on the dual frequency comb technique \cite{giorgetta_2013}, has also been proposed for use in space-based distributed coherent sensing \cite{flood_2025}.
However, high precision time transfer between a ground and space terminal has yet to be demonstrated, and the characteristics of the optical link (through Earth's atmosphere, and to a potentially rapidly moving platform, depending on the orbit) pose significant technical challenges.
RF frequency references based on optical atomic clocks and cavity-stabilized lasers, with their high stability over short timescales, have significant potential but must be qualified for operation in the space environment \cite{yi_2020}. 
Separate efforts have developed and are demonstrating a molecular iodine optical clock and a strontium optical lattice clock for space \cite{doringshoff_2019, guo_2021}. 
However, the reported performance of the former barely meets the stability requirements for future sVLBI observation frequencies ($\geq$ \SI{345}{\giga\Hz}), and the complexity of the latter makes it ill-suited to serve as a frequency reference for another dedicated instrument. 

In this effort, we focus on developing a cavity-stabilized laser-based frequency reference for sVLBI missions.
A fully-stabilized optical frequency comb is employed to transfer the stability of the laser to a \SI{100}{\mega\Hz} RF signal via a technique referred to as optical frequency division (OFD) \cite{fortier_2011}.
While compact cavity-stabilized lasers for space have been developed and demonstrated as part of the Gravity Recovery and Climate Experiment Follow-On (GRACE-FO) mission \cite{abich_2019}, cavity-stabilized laser and frequency comb-based OFD systems have been largely limited to terrestrial demonstrations of synthesizing microwaves with ultra-low phase noise \cite{karlen_2021, giunta_2020}.
We leverage the development of an ultra-stable laser for the Laser Interferometer Space Antenna (LISA) mission and commercial efforts towards a robust, space-qualifiable optical frequency comb to demonstrate a high precision frequency reference with a clear path towards qualification for spaceflight.
The implementation and experimental results of this system are presented, and we discuss the performance of this system when used as a frequency reference for VLBI. 

\begin{figure*}
\centering\includegraphics[width=\linewidth]{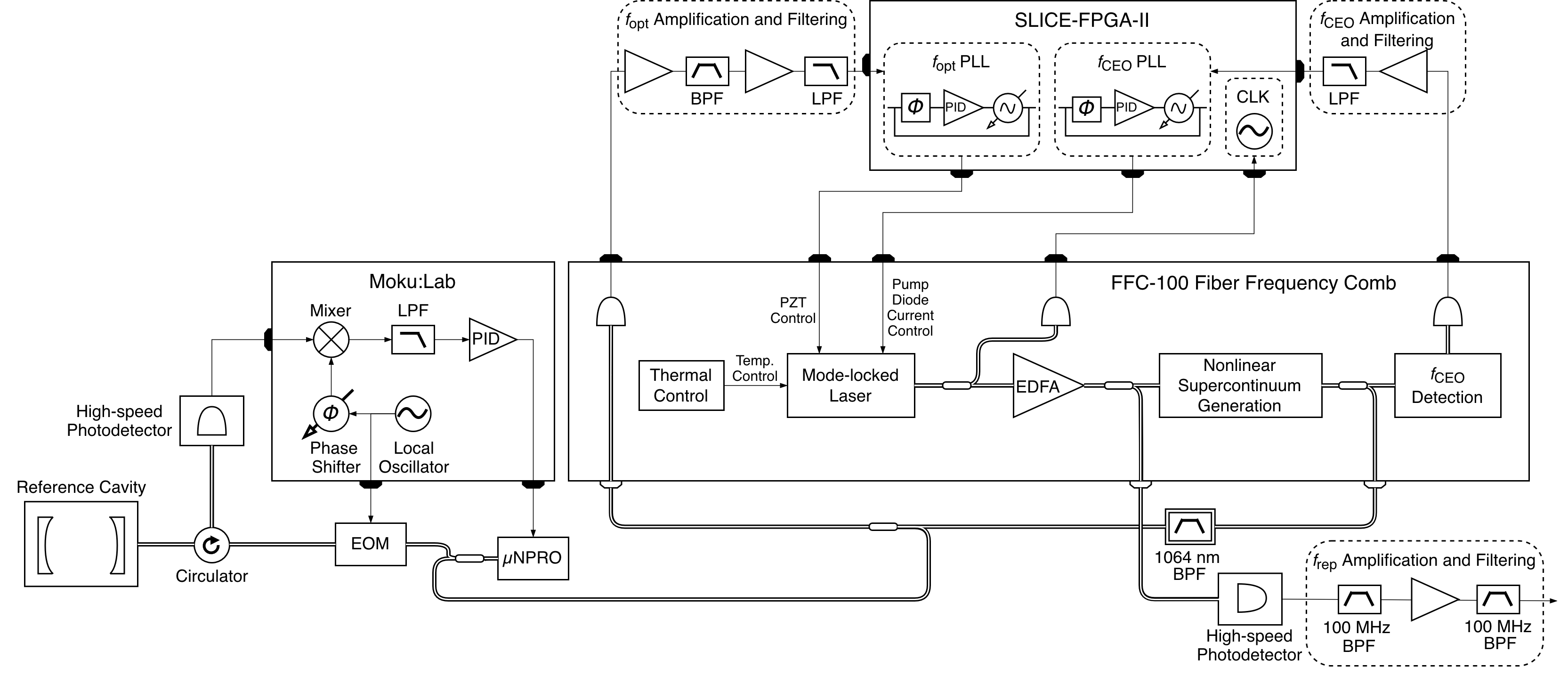}
\caption{\label{fig:single_setup} Experimental setup to photonically generate a \SI{100}{\mega\Hz} frequency reference based on a cavity-stabilized laser and optical frequency comb. A low noise NPRO-based seed laser (µNPRO) is stabilized to a high finesse optical cavity, and this light is in turn used to stabilize the repetition rate $f_\mathrm{rep}$ of a self-referenced fiber frequency comb. The \SI{100}{\mega\Hz} reference signal is generated by detecting, filtering, and amplifying the comb repetition rate.}
\end{figure*}

\section{Experimental Setup}
The experimental setup used to photonically generate the RF reference signal is shown in Figure~\ref{fig:single_setup}.
The cavity-stabilized laser was implemented by stabilizing a prototype of the LISA local oscillator with a prototype of the high finesse optical cavity from the LISA Frequency Reference System (FRS).  
The LISA local oscillator is based on the original monolithic non-planar ring oscillator (NPRO) design from \cite{kane_1985}, but it utilizes a scaled-down crystal to achieve a smaller form factor and is thus referred to as a “micro” NPRO (µNPRO).
The LISA optical reference cavity borrows its design from the optical cavity used in the Laser Ranging Interferometer (LRI) payload of the GRACE-FO mission, with improvements made to the cavity enclosure \cite{yu_2026}.
The thermally stabilized reference cavity is made of an ultra low expansion (ULE) glass spacer with a length of \SI{77.5}{\milli\meter} \cite{thompson_2011}. 
The cavity-stabilized µNPRO for LISA is compliant with the LISA mission requirement on laser frequency noise, namely \SI{30}{\Hz\per\sqrt{\Hz}}$\cdot\sqrt{1+(\SI{2}{\milli\Hz}/f)^4}$ for \SI{0.1}{\milli\Hz} to \SI{1}{\Hz}~\cite{yu_2026}.
The fractional frequency stability is given by dividing this requirement by the central frequency of the laser (\SI{281.627}{\tera\Hz}).
A small fraction of the µNPRO outgoing beam is picked off to the optical cavity, and the laser frequency is locked to the stability of the cavity length via the Pound-Drever-Hall (PDH) technique, implemented using a digital servo (Liquid Instruments Moku:Lab) with a lock bandwidth of \SI{\sim 30}{\kilo\Hz} \cite{drever_1983}.

A self-referenced fiber frequency comb (Vescent Technologies FFC-100) with a repetition rate ($f_\mathrm{rep}$) of \SI{100}{\mega\Hz} was phase-locked to the cavity-stabilized laser. 
To stabilize the carrier-envelope phase of the comb, the internal mode-locked laser was first amplified with an erbium-doped fiber amplifier (EDFA) and spectrally broadened through supercontinuum (SC) generation in a nonlinear medium. 
The carrier-envelope offset frequency ($f_\mathrm{CEO}$) was detected via an interferometric self-referencing scheme and then stabilized with a proportional-integral-derivative (PID) control loop that acts on the oscillator pump diode current. 
This control loop was implemented in a separate field programmable gate array (FPGA) controller (Vescent Technologies SLICE-II), with a lock bandwidth of \SI{\sim 148}{\kilo\Hz}. 

The \SI{100}{\mega\Hz} pulse repetition rate was stabilized by generating a heterodyne beat signal between the \SI{1064}{\nano\meter} cavity-stabilized laser and the broadened spectrum of the \SI{1550}{\nano\meter}-centered comb, which was first filtered through a narrow optical bandpass filter (\SI{2}{\nano\meter} bandwidth) to isolate the comb spectrum around \SI{1064}{\nano\meter}. 
The optical beatnote was detected and its frequency ($f_\mathrm{opt}$) was stabilized via another PID control loop that varies the cavity length using a voltage-modulated intra-cavity piezoelectric transducer (PZT). 
This $f_\mathrm{opt}$ lock featured a lock bandwidth of \SI{\sim 67}{\kilo\Hz}.
Slow cavity length actuation was also conducted by changing the cavity temperature with a thermal controller and was used to maintain the repetition rate stabilization long term. 

\begin{figure}
\centering\includegraphics[width=\columnwidth]{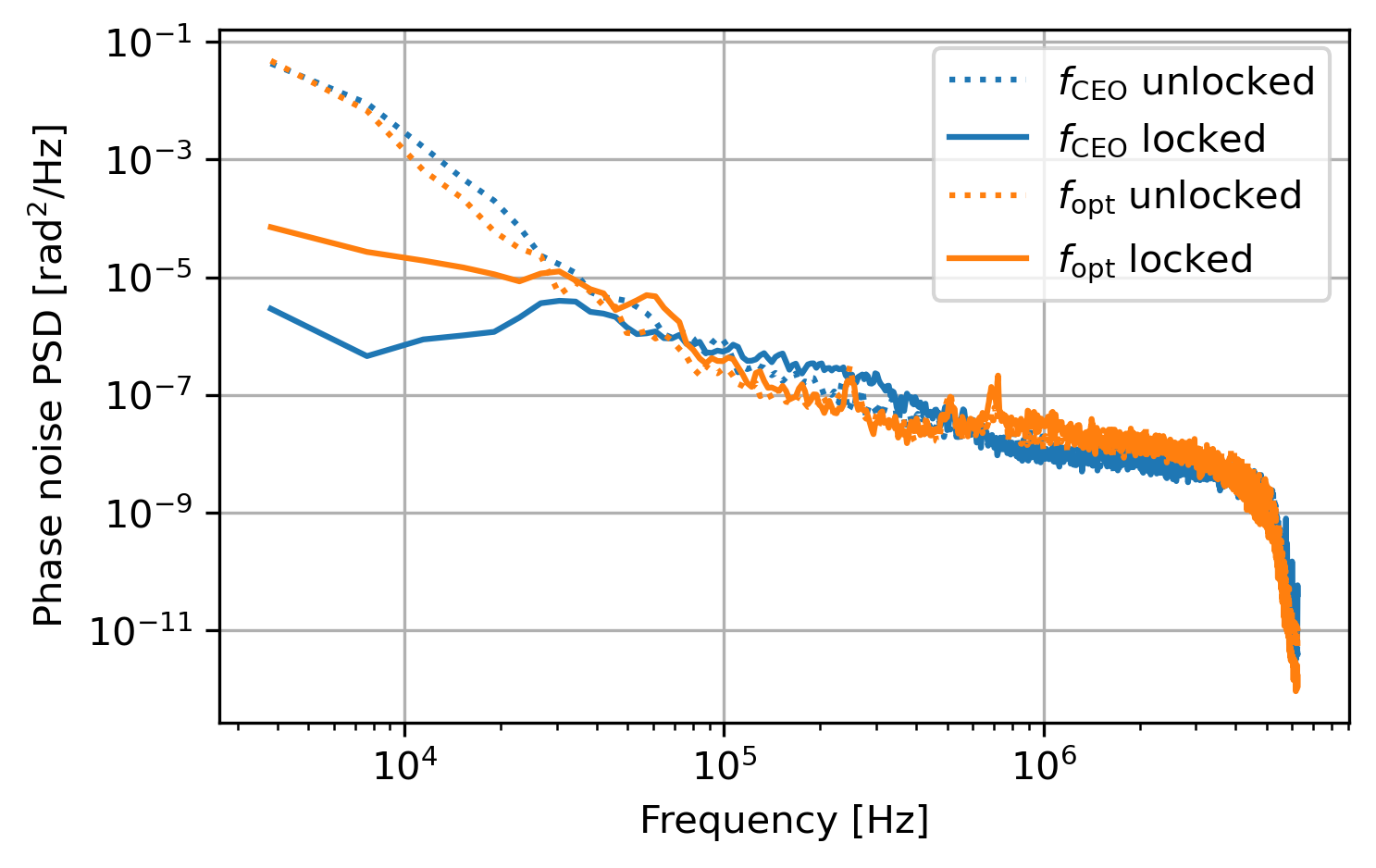}
\caption{\label{fig:comb_lock} Phase noise power spectral densities (PSDs) of the optical frequency comb's $f_\mathrm{CEO}$ and $f_\mathrm{opt}$ beatnotes, when unlocked and locked, showing the suppression of phase noise at lower frequencies by the beatnote stabilization. The integrated phase noise of the stabilized $f_\mathrm{CEO}$ beatnote is \SI{255}{\milli\radian} and of the stabilized $f_\mathrm{opt}$ beatnote is \SI{256}{\milli\radian} (both measured over \SI{3.8}{\kilo\Hz} to \SI{6.2}{\mega\Hz}), indicating that the comb can be fully stabilized.}
\end{figure}

The performance of the frequency comb stabilization is represented by the measured phase noise power spectral densities (PSDs) of the $f_\mathrm{CEO}$ and $f_\mathrm{opt}$ beatnotes, shown in Figure~\ref{fig:comb_lock}. 
The integrated phase noise for the $f_\mathrm{CEO}$ lock is \SI{255}{\milli\radian}, measured over \SI{3.8}{\kilo\Hz} to \SI{6.2}{\mega\Hz}, and similarly the integrated phase noise for $f_\mathrm{opt}$ is \SI{256}{\milli\radian}. 

The stabilized repetition rate was detected internally and used to provide a clock signal to the FPGA that implements both control loops.
For the RF reference signal used in the stability measurement, $f_\mathrm{rep}$ was detected after the internal EDFA of the FFC-100 using an external high-speed photodetector (Coherent ET-3500F) that featured a lower noise floor. 
The first harmonic was selected from the resulting RF signal with a narrow bandpass filter, and was then amplified to obtain the desired \SI{100}{\mega\Hz} frequency reference. 
This output reference frequency was selected to align with those of time and frequency systems typically employed in spacecraft systems, such as conventional oven controlled crystal oscillators (OCXOs) and ultra-stable oscillators (USOs).
These oscillators typically operate at MHz radio frequencies \cite{rakon_2025, quanticwenzel_2025}.

\begin{figure}
\centering\includegraphics[width=\columnwidth]{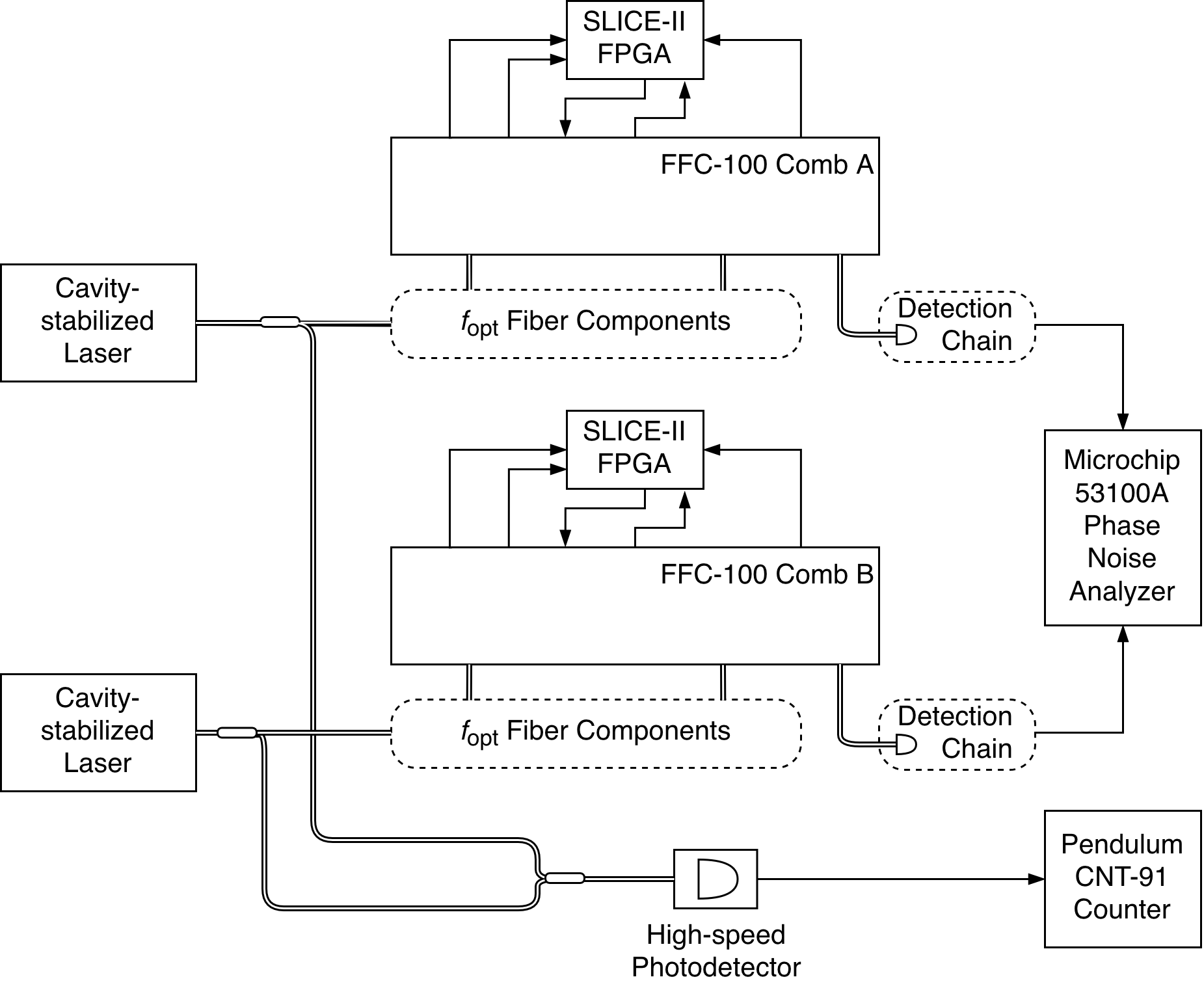}
\caption{\label{fig:full_setup} Setup to measure the performance of the comb and cavity-stabilized laser-based frequency reference. As the phase stability of a single reference is better than that of most measurement instruments, a second, identical system was implemented and used to characterize the overall performance. The two cavity-stabilized lasers were simultaneously compared to track the performance of the optical references. }
\end{figure}

As the performance of this system was higher than that of most readily available frequency references, a second, identical system was implemented to characterize the stability of the output RF reference signal. 
This secondary setup featured another FFC-100 frequency comb locked to another cavity-stabilized laser consisting of a separate µNPRO and high finesse optical cavity. 
The relative stabilities of the \SI{100}{\mega\Hz} reference signals were compared using a Microchip 53100A Phase Noise Analyzer. 
At the same time, the two cavity-stabilized lasers were compared directly to determine the stability of the optical references.
Light was picked off from both lasers and directly mixed on a fast photodetector (Thorlabs RXM42AF). 
The instantaneous frequency of the resulting beat signal was measured with a Pendulum CNT-91 counter, and used to derive the stability measurements.
This setup is depicted in Figure~\ref{fig:full_setup}. 

\section{Performance and Characterization}
The performance was assessed in terms of the Allan deviation between the two RF reference signals.
The measurements shown here are calculated as the overlapping Allan deviation to provide better confidence, especially at longer averaging times, and to align with the VLBI frequency stability requirements. 
As the two comb and cavity-stabilized laser setups are identical, we can compare the two separate systems and analyze their performance with the assumption that both contribute uniformly to the phase noise.
Therefore, a measurement obtained by comparing the two setups can be divided by $\sqrt{2}$, assuming incoherent noises from both, to represent the performance of a single system. 
The measurements throughout this section are differential between two instances of the same kind of device; hence, we apply the by-$\sqrt(2)$ division to convert all such measurements to a single-instance noise contribution.

\begin{figure}
\centering\includegraphics[width=\columnwidth]{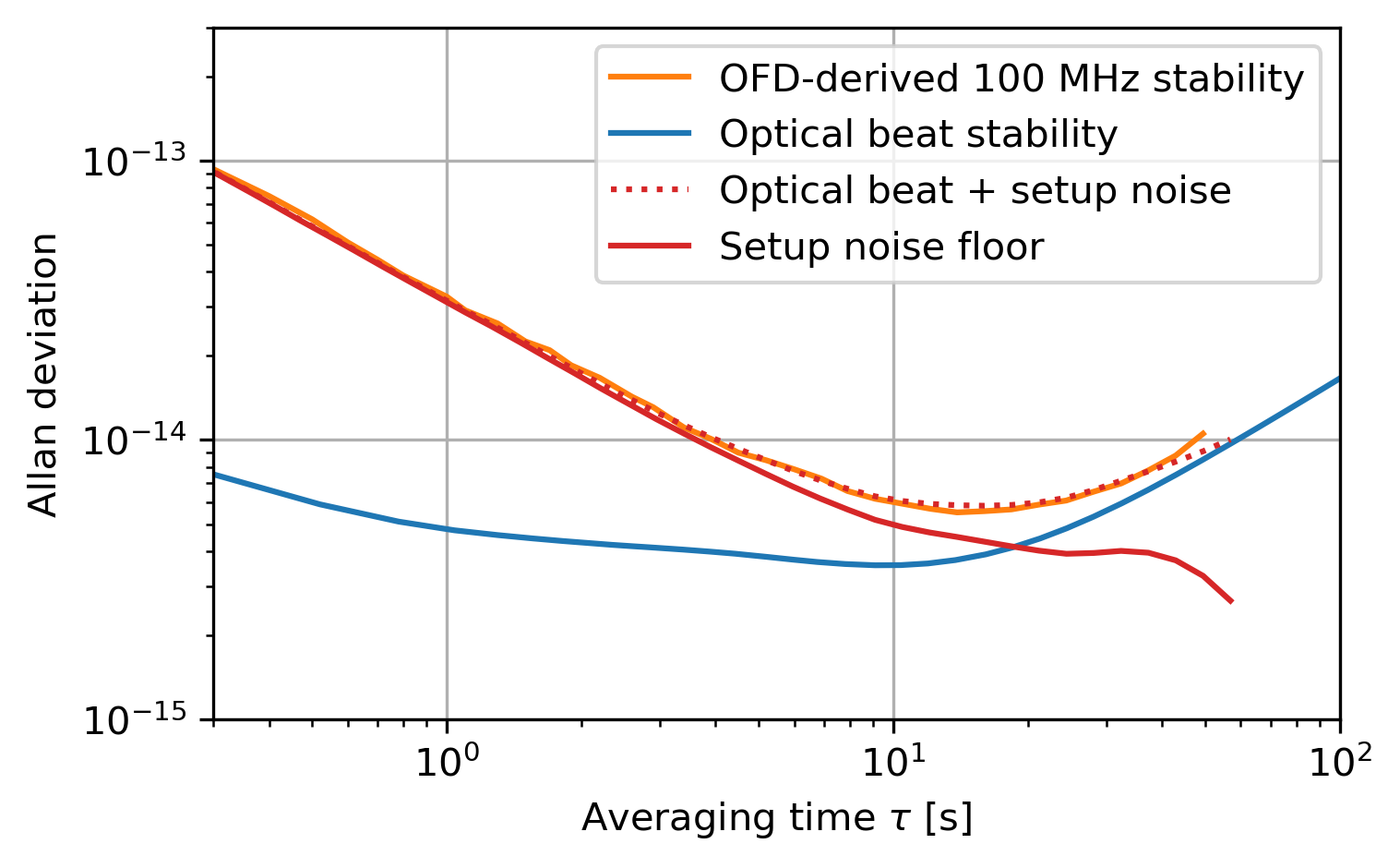}
\caption{\label{fig:performance} Allan deviation of a single stabilized \SI{100}{\mega\Hz} reference as compared to the optical stability of the cavity-stabilized reference laser and the OFD setup noise floor, showing the limitation from the experimental setup noise.}
\end{figure}

The Allan deviation of the single \SI{100}{\mega\Hz} reference is presented as a function of the averaging time and displayed as the orange trace in Figure~\ref{fig:performance}.
For the stability of the OFD-generated \SI{100}{\mega\Hz} signal, Allan deviations of \num{3.26d-14} at \SI{1}{\second}, \num{6.00d-15} at \SI{10}{\second}, and \num{6.73d-15} at \SI{30}{\second} were obtained.
This performance approaches that of the optical stability of the cavity-stabilized reference laser, which is derived from the heterodyne beatnote between the two cavity-stabilized lasers and is shown in blue in Fig.~\ref{fig:performance}, after \SI{10}{\second}.
The optical reference achieves Allan deviations of \num{4.75d-15} at \SI{1}{\second}, \num{3.57d-15} at \SI{10}{\second}, and \num{5.75d-15} at \SI{30}{\second}.
The \SI{100}{\mega\Hz} RF reference performance is limited by the measured noise floor (red trace in Fig.~\ref{fig:performance}) of the setup at integration times less than \SI{10}{\second}.
After \SI{10}{\second}, the frequencies of the cavity-stabilized lasers begin to drift. 
The RF reference results match the expected performance (dotted red trace in Fig.~\ref{fig:performance}), which is calculated by adding the instabilities of the optical heterodyne beatnote and the total OFD setup noise.

\begin{figure}
\centering\includegraphics[width=\columnwidth]{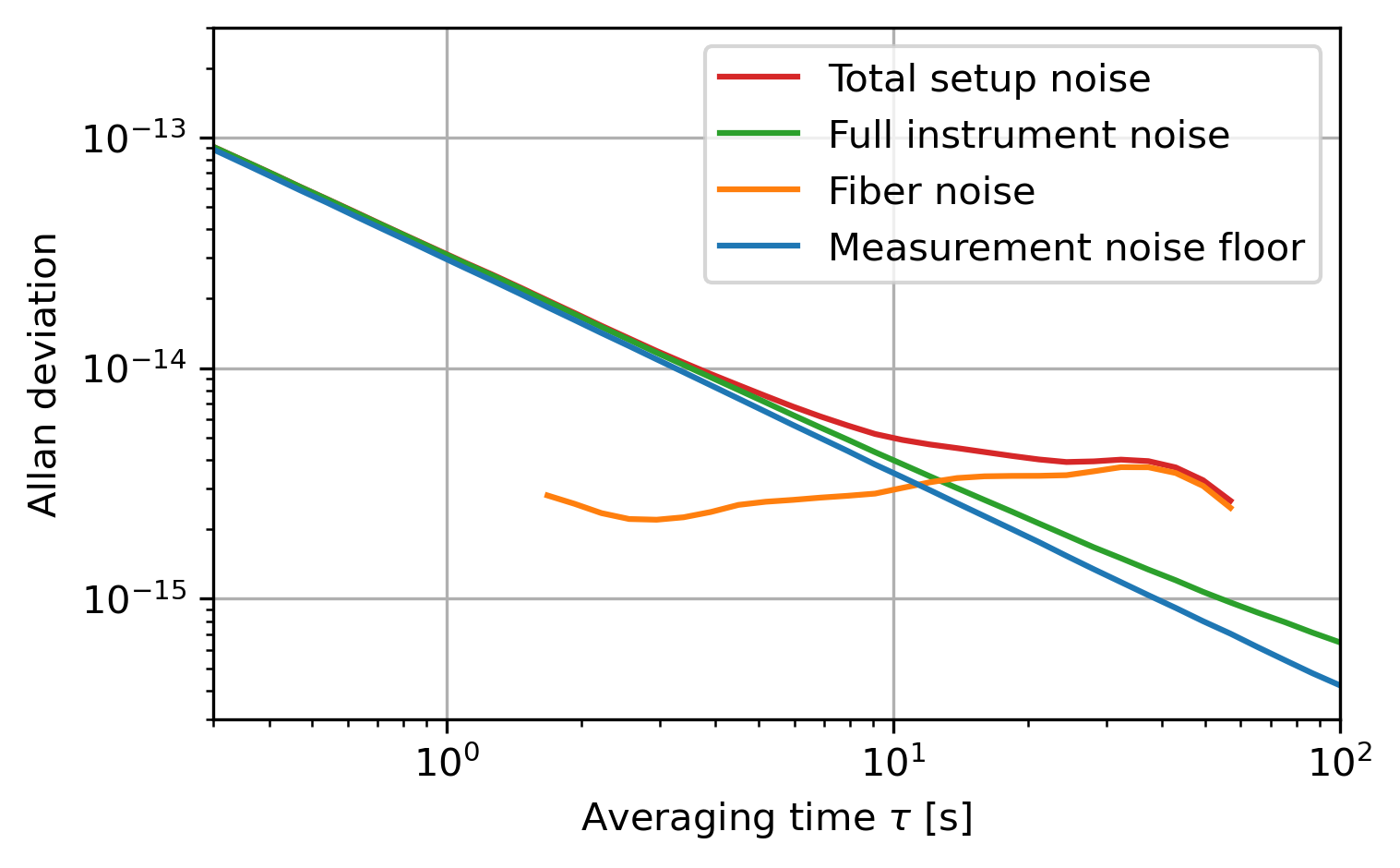}
\caption{\label{fig:noise} Allan deviations representing the total noise floor of a single setup and its primary components: the full instrument noise (encompassing all of the components in an OFD and measurement
setup) and the noise contribution from the long, uncompensated fiber path between a cavity-stabilized laser and the OFD and measurement instrumentation. The main contribution to the full instrument noise, and thus, the total setup at $\tau$\SI{< 10}{\second}, is the noise floor of the Microchip 53100A measurement instrument. For $\tau$\SI{> 10}{\second}, the total setup noise is dominated by the fiber noise.}
\end{figure}

This setup noise floor and its components are presented in Figure~\ref{fig:noise}. 
The total noise (shown in red) was determined by adding in quadrature the two primary contributions: the full instrument noise, which consists of the OFD and measurement setup noise, and the noise resulting from vibration and temperature fluctuations in the fiber optic patch cables that connect the cavity-stabilized reference lasers to this instrumentation. 
The former encompasses all of the components in the OFD and measurement setup, including the frequency comb and comb phase-locking noise contributions, repetition rate detection photodetector, various RF filters and amplifiers, and the measurement equipment.

This noise floor was measured by splitting the light from one cavity-stabilized laser and phase-locking both frequency combs in the two OFD setups to this same reference. 
By measuring the relative instability between the two RF signals derived from the same reference laser, the noise contribution to a single \SI{100}{\mega\Hz} reference from its OFD and measurement setup was recorded. 
This is shown in green Fig.~\ref{fig:noise}. 
This noise level limits the overall performance at \SI{1}{\second}. 
However, the dominant factor at short averaging times was found to be the noise floor of the measurement instrument, the Microchip 53100A Phase Noise Analyzer, which is represented by the blue trace in Fig.~\ref{fig:noise}. 
To measure this performance limit, a \SI{100}{\mega\Hz} reference output from one of the OFD setups was split and fed to both inputs of the test device. 
The resulting stability measurement thereby describes the noise floor of this instrument at \SI{100}{\mega\Hz}.

Another significant source of noise at averaging times \SI{>10}{\second} was discovered to be the long fiber optic cables that connect the cavity-stabilized lasers to the OFD and measurement setup.
As these systems are set up on separate optical tables, the length of this non-fiber noise canceled path is approximately \SI{10}{\meter}.
In order to account for this noise contribution, measurements of the cavity-stabilized lasers’ optical heterodyne beatnote were taken before and after this fiber path. 
The Allan deviation was calculated, and the RMS difference between the “after” and “before” stabilities is shown in orange in Fig.~\ref{fig:noise}.
This shows that while for $\tau$\SI{< 10}{\second} the total setup noise floor is dictated by the noise of the Microchip 53100A measurement instrument, the long fiber path also limits the Allan deviation for $\tau >$\SI{10}{\second}.
However, at these time scales, the performance is already limited by the cavity drift of the optical references.

\section{Discussion}
While the frequency stability measurements are limited by uncompensated fiber paths and the noise floor of our measurement instrument, these results still demonstrate that this frequency reference is relevant for future sVLBI missions.
Beyond the performance bound presented in Eq.~\ref{eqn:basic_coherence_func}, the coherence time can be calculated more rigorously and defined in terms of a coherence function $C(T)$:
\begin{equation}
    C(T) = \bigg | \frac{1}{T} \int^T_0 \exp[i\phi(t)] dt \bigg |.
\end{equation}
This $C(T)$ is proportional to the amplitude of the interference fringes, as $\phi(t)$ is the difference in phase between two stations forming an interferometer and $T$ is the integration time \cite{rogers_1981}. 
The root-mean-square (RMS) value of $C(T)$, also referred to as the “coherence,” decreases monotonically with time, with resulting values ranging from 1 to 0, such that when $\langle C^2(T)\rangle^{1/2} = 1$ there is no loss of coherence.
The coherence time can then be defined as the value of $T$ for which $\langle C^2(T)\rangle^{1/2}$ drops to a specified value below 1 \cite{thompson_2017}. 
The coherence $\langle C^2(T)\rangle^{1/2}$ is a commonly used figure-of-merit for VLBI frequency standards \cite{nand_2011}. 

The $\phi(t)$ in its calculation includes the effects of frequency reference instability, atmospheric fluctuations, and other sources of phase noise.
However, if only the phase difference due to the stability of the frequency reference is considered, the coherence can be used to determine a performance requirement for the reference.
The mean-squared value of $C(T)$ can be expressed in terms of the Allan variance of the reference, $\sigma^2_y$,
\begin{equation}
    \langle C^2(T) \rangle = \frac{2}{T} \int^T_0 (1 - \frac{\tau}{T}) \exp\{-\pi^2 \nu^2_0 \tau^2 [\sigma^2_y(\tau) + \sigma^2_y(2\tau) + \dots]\} d\tau,
    \label{eqn:coherence_func} 
\end{equation}
where $\nu_0$ is the observing frequency (in \unit{\Hz}) and $\sigma_y^2(\tau)$ is the Allan variance at an integration time of $\tau$ \cite{thompson_2017}. 
This calculation assumes that the same performance reference is used at each telescope in the pair.

\begin{figure}
\centering\includegraphics[width=\columnwidth]{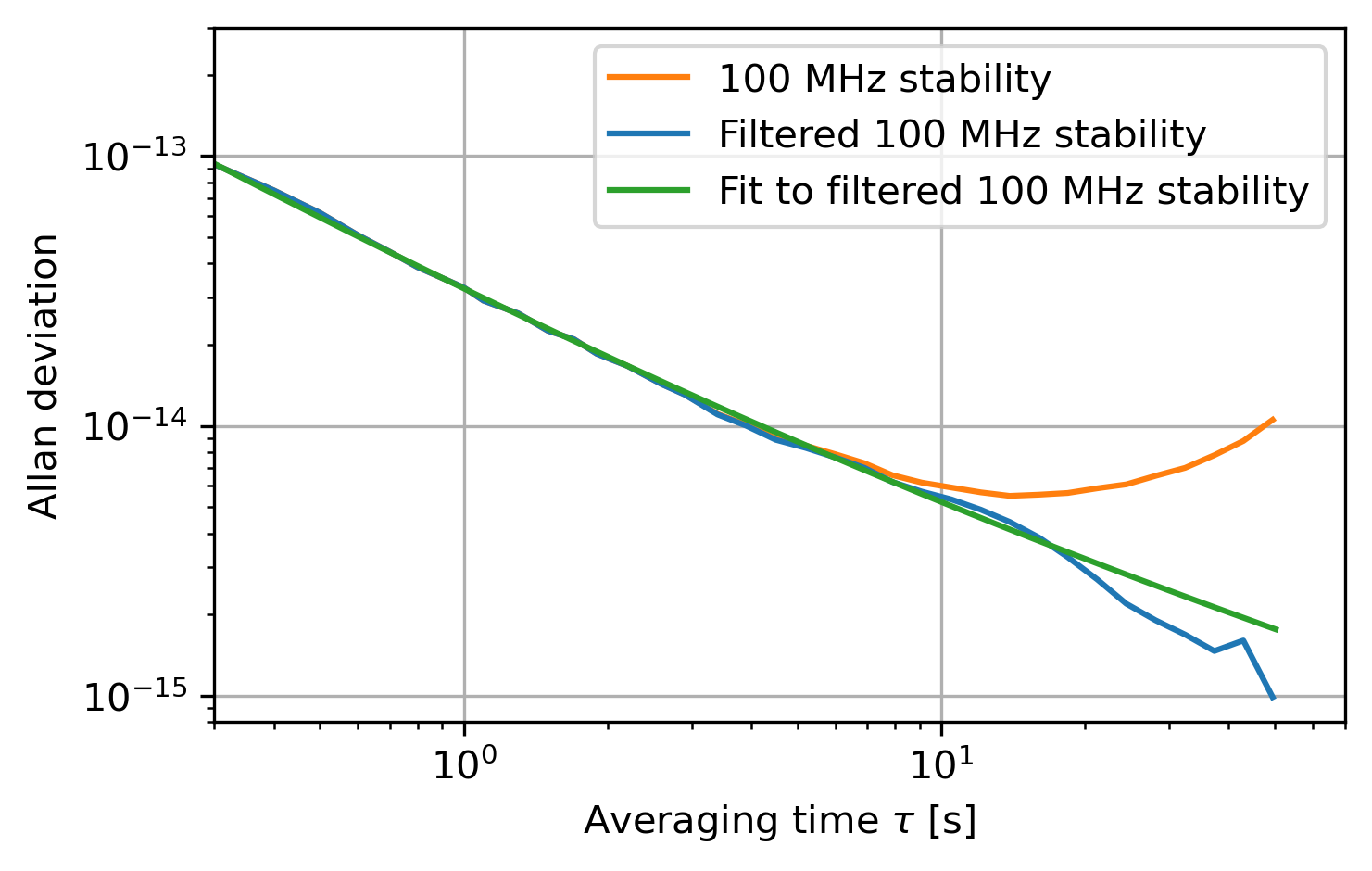}
\caption{\label{fig:perf_filter_fit} Comparison of the photonically-generated \SI{100}{\mega\Hz} signal stability (in blue), its stability after filtering the frequency drift (in orange), and the fit to the filtered result (in green).}
\end{figure}

Based on the measured stability of the frequency reference developed in this work, the coherence for various observing frequencies can be determined using Equation~\ref{eqn:coherence_func}.
However, we first address the frequency drift in the measurement record caused by the drifting central frequencies of the cavity-stabilized lasers by subtracting the quadratic frequency trend.
Then, to obtain an analytical expression for $\sigma_y(\tau)$ from the measured stability, a non-linear least squares fit to the data was taken using a model function of the form $\sigma_y(\tau) = k_1\tau^{-1}+k_2\tau^{-1/2}+k_3\tau^0+k_4\tau^{1/2}+k_5\tau^1$.
The resulting $k_n$ parameters, constrained to be $\geq0$, were $k_1 =$ \num{2.31d-14}, $k_2 =$ \num{9.22d-15}, $k_3 = 0$, $k_4 = 0$, and $k_5 = 0$.
With this expression for $\sigma_y(\tau)$, the coherence can be calculated numerically. 
A comparison of the OFD-generated \SI{100}{\mega\Hz} signal stability, its stability after filtering the quadratic frequency drift, and the fit to this experimental result is presented in Figure~\ref{fig:perf_filter_fit}.

\begin{figure}
\centering\includegraphics[width=\columnwidth]{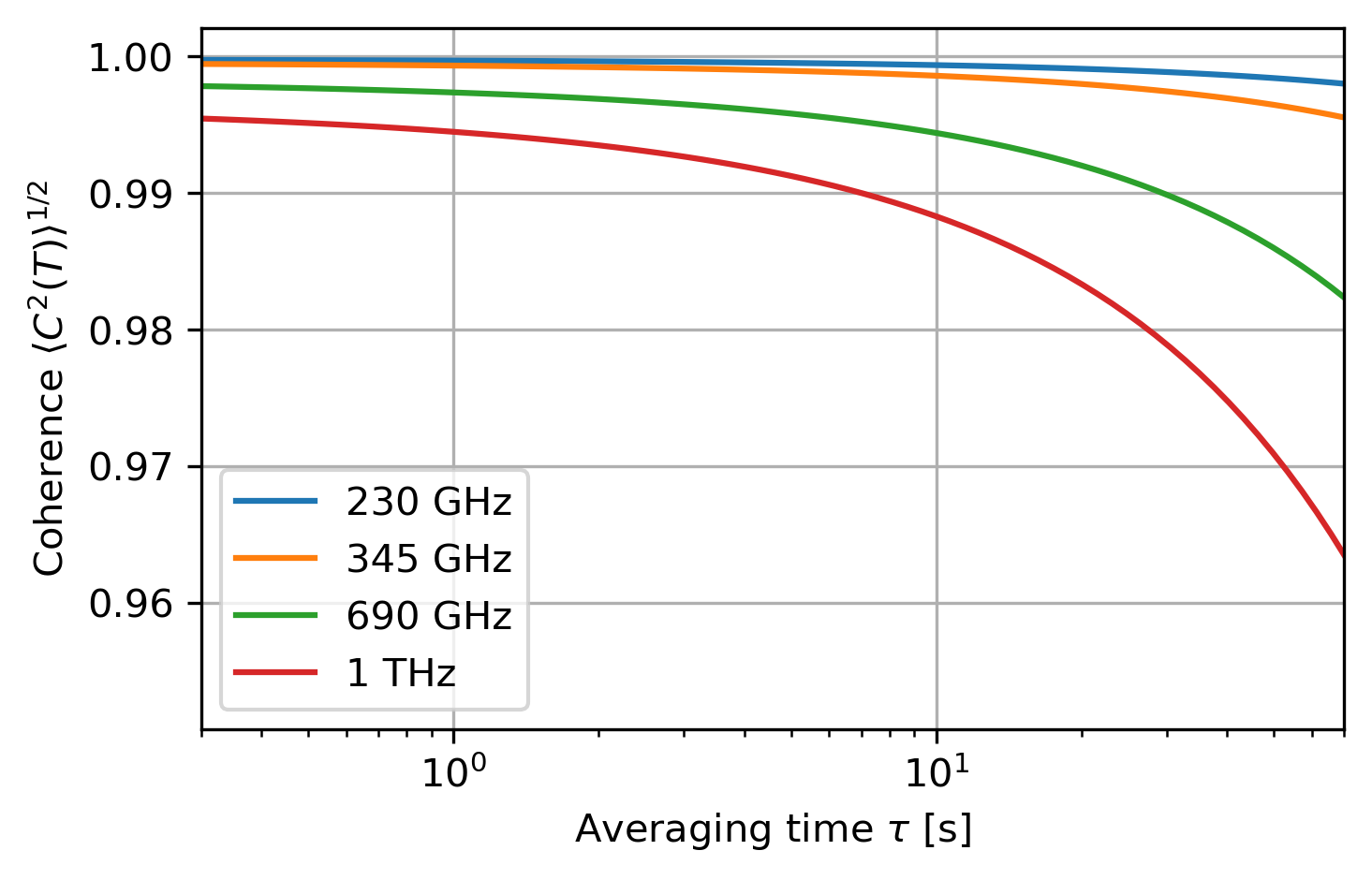}
\caption{\label{fig:coherence} Calculated coherence $\langle C^2(T)\rangle^{1/2}$ for averaging times from \qtyrange[range-units=single,range-phrase=-]{0.1}{100}{\second} and observing frequencies of \SI{230}{\giga\Hz}, \SI{345}{\giga\Hz}, \SI{690}{\giga\Hz}, and \SI{1}{\tera\Hz} using the frequency reference developed in this effort. }
\end{figure}

As the filtered \SI{100}{\mega\Hz} reference is dominated by white phase and white frequency noise, the sum of the Allan variances $\sum_{i=0}^n \sigma^2_y(2^i\tau)$ in Eqn.~\ref{eqn:coherence_func} converges, theoretically, as $n \xrightarrow{} \infty$.
More practically, in this implementation, the coherence estimates converge within 10 iterations. 
We calculate the coherence values using this frequency reference for observing frequencies of \SI{230}{\giga\Hz}, \SI{345}{\giga\Hz}, \SI{690}{\giga\Hz}, and \SI{1}{\tera\Hz}.
These results are shown in Figure~\ref{fig:coherence}. 
These show that coherences greater than 0.95 can be maintained for integration times of up to \SI{100}{\second} for all observing frequencies, even \SI{1}{\tera\Hz}. 
Coherence values greater than 0.8 or 0.9 are common benchmarks for VLBI \cite{rogers_1981, nand_2011, zhao_2025}. 
Thus, this frequency reference can meet the needs of future sVLBI concepts. 

Furthermore, a clear path to qualification for operation in the space environment exists for this system, as its primary components are separately being developed and evaluated for operation in space. 
The cavity-stabilized reference laser is borrowed from ongoing development efforts for the LISA mission.  
The µNPRO is a critical part of the LISA laser optical module, and a robust version of this module intended for operation in the relevant space environment has been developed. 
The optical reference cavity already has demonstrated space heritage through the GRACE-FO mission.
At the time of writing, this system is undergoing rigorous environmental testing and its performance will continue to be evaluated in anticipation of the LISA launch date in 2035 \cite{yu_2026}.
A ruggedized, modular version of the optical frequency comb from Vescent Technologies has also been developed. 
This comb module is radiation-hardened-by-design has been subject to TVAC and vibration and shock testing \cite{timmers_2024}.

\section{Conclusion}
Highly stable frequency references, with a particular emphasis on short-term stability, are necessary for the next generation of sVLBI missions, which target higher observation frequencies than current VLBI techniques. 
In this work, we present a frequency reference that meets these needs based on cavity-stabilized laser development for the LISA mission and commercial efforts towards a robust, space-qualifiable optical frequency comb.
This system achieves an Allan deviation of \num{3.26d-14} at \SI{1}{\second}, \num{6.00d-15} at \SI{10}{\second}, and \num{6.73d-15} at \SI{30}{\second}, a performance limited by the noise floor of the measurement instrument, uncompensated fiber paths, and beyond \SI{30}{\second} the drift of the cavity-stabilized laser frequencies. 
Nonetheless, we demonstrate that this performance can meet the needs of future sVLBI missions with observation frequencies up to \SI{1}{\tera\Hz}, maintaining a coherence greater than 0.95 for the relevant integration times. 

Future work will address the uncompensated fiber lengths and pursue alternative measurement approaches.  
We also seek to fully qualify the optical frequency comb for operation in the space environment.
Ultimately, this effort is an initial step towards the implementation of a highly stable, fully space-qualified frequency reference for future sVLBI missions.

\begin{acknowledgments}
This effort has been supported by the Internal Research and Development (IRAD) program at NASA Goddard Space Flight Center.
K.Y.’s work is supported by NASA under Award No.80GSFC24M0006.
H.T.'s work is supported by a NASA Space Technology Graduate Research Opportunity, grant no. 80NSSC21K1277. 
H.L. and C.Z.’s work is supported by the NASA Space Technology Mission Directorate Early Career Initiative Award. 
C.Z.’s work is supported in part through the Arizona NASA Space Grant Consortium, Cooperative Agreement 80NSSC25M7084.

\end{acknowledgments}

\bibliography{final}

\end{document}